\documentclass[aps,pra,amsmath,amssymbs,superscriptaddress]{revtex4}
\usepackage{times}
\usepackage{graphicx}
\usepackage{epsfig}
\usepackage{wrapfig}
\usepackage{bbm}
\usepackage[usenames]{color}
\usepackage{array}
\usepackage{setspace}

\newcommand{\be}{\begin{equation}}
\newcommand{\ee}{\end{equation}}

\def\Kex{\kappa_{ex}}
\def\Kin{\kappa_{in}}
\def\J{J}
\def\Phii{\phi}
\def\Lambda0{\lambda_{0}}
\def\W0{\omega_{0}}
\def\V0{\nu_{0}}

\def\dW0{\Delta\omega_{0}}
\def\dV0{\Delta\nu_{0}}
\def\dLambda0{\Delta\lambda_{0}}
\def\dPhii{\Delta\phi}

\begin{document}

\title{Topologically Robust Transport of Photons in a Synthetic Gauge Field}
\author{S. Mittal} 
\affiliation{Joint Quantum Institute, NIST/University of Maryland, College Park MD 20742, USA}
\author{J. Fan}
\affiliation{Joint Quantum Institute, NIST/University of Maryland, College Park MD 20742, USA}
 \author{S. Faez} 
 \affiliation{Huygens-Kamerlingh Onnes Laboratorium, Universiteit Leiden, Postbus 9504, 2300 RA Leiden, The Netherlands}
 \author{A. Migdall}
 \affiliation{Joint Quantum Institute, NIST/University of Maryland, College Park MD 20742, USA}
 \author{J. M. Taylor}
 \affiliation{Joint Quantum Institute, NIST/University of Maryland, College Park MD 20742, USA}
 \author{M. Hafezi}
 \email{hafezi@umd.edu}
\affiliation{Joint Quantum Institute, NIST/University of Maryland, College Park MD 20742, USA}

\begin{abstract}
 Electronic transport in low dimensions  through a disordered medium leads to localization. The addition of gauge fields to disordered media leads to fundamental changes in the transport properties. For example, chiral edge states can emerge in two-dimensional systems with a perpendicular magnetic field. Here, we implement a ``synthetic'' gauge field for photons using silicon-on-insulator technology. By determining the distribution of transport properties, we confirm the localized transport in the bulk and the suppression of localization in edge states, using the ``gold standard''  for localization studies.  Our system provides a new platform to investigate transport properties in the presence of synthetic gauge fields, which is important both from the fundamental perspective of studying photonic transport and for applications in classical and quantum information processing.
\end{abstract}
\maketitle
\setstretch{1.2}
%%%%%%%%%%%%%%%%%%%%%%%%%%%%%%%%%%%%%%%%%%%%%%%%%%%%%%%%%%%%%%%%%%%%%%%%%%%%%%%%%%%%%%%%%%%%%%%%%%%%%%%%%%%%%%%%%%%%%%%%%%%%
%  Introduction -
%%%%%%%%%%%%%%%%%%%%%%%%%%%%%%%%%%%%%%%%%%%%%%%%%%%%%%%%%%%%%%%%%%%%%%%%%%%%%%%%%%%%%%%%%%%%%%%%%%%%%%%%%%%%%%%%%%%%%%%%%%%%
Photons provide a natural and convenient medium to investigate fundamental quantum transport properties \cite{Segev:2013,Lagendijk:2009}. Using photons, one can selectively excite states, and observe both spectral and spatial response throughout the material, which are challenging tasks in electronic systems. However, experimental efforts studying gauge fields with photons, have been limited to the microwave domain \cite{Raghu:2008,Wang:2009}, while investigations in the optical domain have remained elusive. This is due to the fact that magneto-optic effects --- the simplest source of coupling between gauge fields and photons ---  are extremely weak at optical frequencies. Recently though, there have been a significant number of proposals  to synthesize gauge fields for optical photons \cite{Hafezi:2011,Umucalilar:2011,Fang:2012,Kraus:2012,Rechtsman:2012,Khanikaev:2013}. In particular, two concurrent experiments showed exemplary signatures of topological edge states through direct imaging \cite{Hafezi:2013,Rechtsman:2013}.  Here we report the first observation of the robust nature of topologically-protected edge states using an analysis of the statistics of transport properties (transmission and delay). We use a 2D lattice of coupled ring resonators with a synthetic magnetic field, implemented using silicon-on-insulator technology. By considering the distribution of Wigner delay times \cite{Chabanov:2000,Chabanov:2001}, we can unambiguously distinguish non-localized diffusive transport in lossy edge states from tunneling through localized bulk states. Finally, we compare the transmission of topologically ordered edge states against the transmission in a topologically trivial one-dimensional system.

%%%%%%%%%%%%%%%%%%%%%%%%%%%%%%%%%%%%%%%%%%%%%%%%%%%%%%%%%%%%%%%%%%%%%%%%%%%%%%%%%%%%%%%%%%%%%%%%%%%%%%%%%%%%%%%%%%%%%%%%%%%%
%  Experiment
%%%%%%%%%%%%%%%%%%%%%%%%%%%%%%%%%%%%%%%%%%%%%%%%%%%%%%%%%%%%%%%%%%%%%%%%%%%%%%%%%%%%%%%%%%%%%%%%%%%%%%%%%%%%%%%%%%%%%%%%%%%%

Our experiments are performed on a two dimensional lattice of coupled ring resonators \cite{Cooper:2010} (Fig.~1a). The ring resonators are coupled using another set of link rings which are designed to be anti-resonant to the main ring resonators, i.e., the length of the connecting rings is slightly longer than main rings so as to acquire an extra $\pi$ phase shift. The link resonators are spatially shifted, along the y-axis,  with respect to the main lattice-site resonators such that transiting photons acquire a phase $y \phi$ when hopping along the x-axis at a lattice site with row index $y$ \cite{Hafezi:2011}. Therefore a round trip along any plaquette (consisting of 4 ring and 4 link resonators, see Fig. 1a) results in a total accumulated phase of magnitude $\phi$ with a $\pm$ sign corresponding to the direction (clockwise or counter-clockwise) of travel along the plaquette. Here we only excite and measure the counter-clockwise mode in the main ring resonators, with the input port as indicated in Fig.~1a. This system is equivalent to a uniform synthetic magnetic field with flux $\phi$ penetrating each plaquette of a 2D photon gas, with the tight-binding Hamiltonian

\begin{equation}
  H_{0} = -\J \sum_{x,y} \hat{a}_{x+1,y}^{\dag} \hat{a}_{x,y}^{} e^{-i \phi y}
  + \hat{a}_{x,y}^{\dag} \hat{a}_{x+1,y}^{} e^{i \phi y}
  + \hat{a}_{x,y+1}^{\dag} \hat{a}_{x,y}^{} + \hat{a}_{x,y}^{\dag} \hat{a}_{x,y+1}^{},
\end{equation}
where $\J$ is the coupling rate between the on-site rings, and $\hat{a}_{x,y}$ and $\hat{a}^{\dag}_{x,y}$ are the photon annihilation and creation operators at a main resonator site with indices $x,y$.\\

\begin{figure*}
 \centering
 \includegraphics[width=0.6\textwidth]{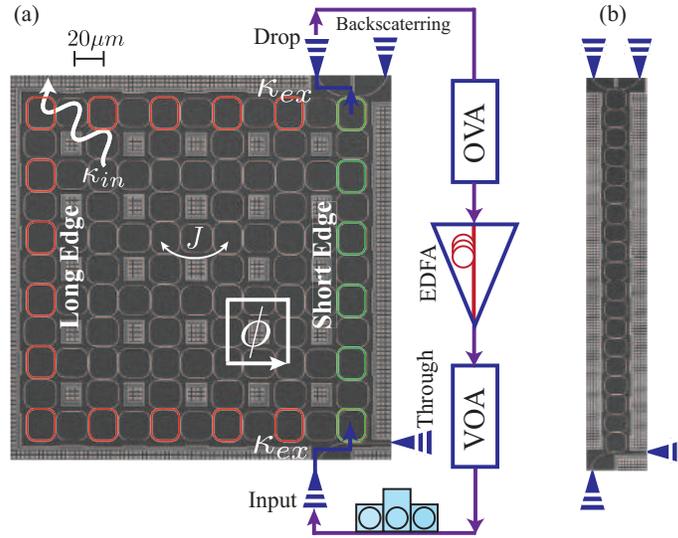}
 \caption{(a) SEM image of a 2D lattice with the measurement setup. Light is coupled into the lattice/chain at input port exciting counterclockwise rotating mode. The output at drop port is measured using an optical vector network analyzer (OVA). Erbium-doped fiber amplifier (EDFA) and variable optical attenuator (VOA) are used to control input power along with a polarization controller. (b) SEM image of a 1D device with 10 main rings. Main rings are coupled using link rings similar to 2D devices.  }
 \label{fig:1}
\end{figure*}

For an infinite lattice, the energy eigenvalues of this Hamiltonian constitute the famous Hofstadter butterfly spectrum. The eigenvalues group into allowed energy bands separated by band-gaps, forming a topological insulator. For a finite lattice, the band-gaps are populated with so-called \emph{edge states}. The edge states are unidirectional, clockwise (or counter-clockwise) propagating states, with their wavefunction confined to the perimeter of the lattice. We call these long-edge-  and short-edge-states (Fig.~\ref{fig:1}a), respectively, because of the length they travel along the lattice edge from input to output port. These states are in sharp distinction with the eigenstates in the allowed energy bands, which are called bulk states because their eigenfunctions occupy resonators in the bulk of the lattice. In the presence of lattice disorders, such as resonance frequency mismatch and coupling variations, bulk states become localized as the lattice size exceeds the localization length \cite{Wen:2007}. For our system, bulk states in even the smallest lattice (4$\times$4) are localized. Edge states, on the other hand, are topologically protected and their wavefunctions are robust against disorder in the lattice. As a result, edge state wavefunctions propagate along the entire edge of the lattice irrespective of the lattice size, although their intensity fall due to absorption and scattering loss. Our goal here is to leverage quantitative measurements of transmission and delay-time to unambiguously demonstrate the robust nature of edge states and distinguish them from bulk states. We show that the edge state transport is diffusive and the delay distribution is gaussian and centered at the average, while for bulk states the delay distribution is asymmetric with the peak value being well below the average, similar to the localization typical in one-dimension \cite{Chabanov:2001,Texier:1999,Xu:2011}.

\begin{figure*}
 \centering
 \includegraphics[width=0.9\textwidth]{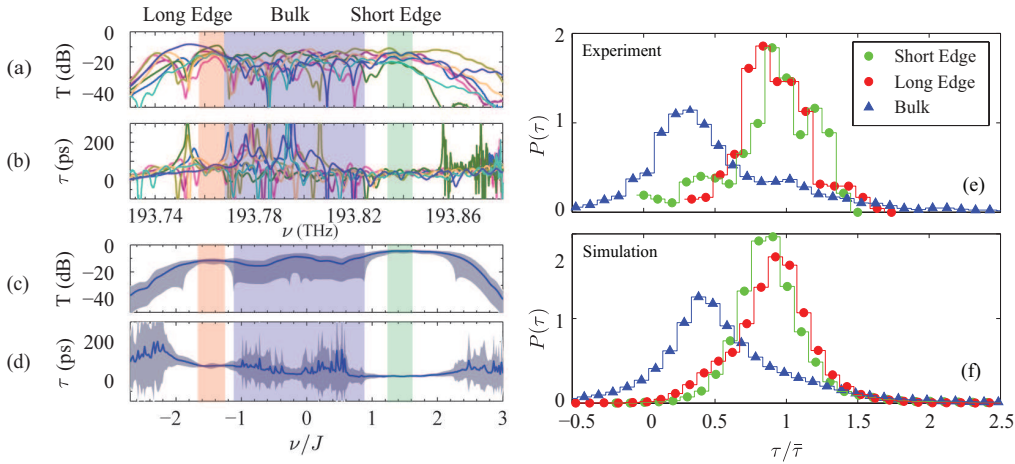}
 \caption{(a) Measured transmission and (b) delay-time spectra for eight 8$\times$8 lattice size devices. The spectra have been normalized and shifted along x-axis to superpose them (see Supplementary). Two regions with reduced variance in transmission and delay are indicated (shaded red and green). Noisy bulk states region is shown in blue. (c,d) Simulated transmission and delay with the average (solid blue line) and 95\% confidence band (grey shaded area) determined from the standard deviation across devices. (e,f) Measured and simulated delay statistics for edge and bulk states. Delay distribution for edge states is gaussian indicating diffusive transport. For bulk states the distribution is asymmetric, showing localized transport. Data is taken across 8 devices. The delays are normalized to the average (rms) and the overall delay distribution is normalized to in-band average and the delay distribution is normalized such that the area under the curve is unity. }
 \label{fig:2}
\end{figure*}

We implemented this system using silicon-on-insulator (SOI) technology, as described in the Supplementary. Fig.~\ref{fig:2}(a,b) shows the observed transmission and delay spectra at the drop port for eight different 8$\times$8 lattice size devices. While the spectra differ significantly because of intrinsic fabrication variations in waveguide dimensions, we can already see the first manifestation of robust edge states in the form of two regions with suppressed variance across devices, in both the transmission and delay spectra (red and green shaded). Since edges states are topologically constrained to travel along the lattice edge, device-to-device fabrication variations in system parameters do not affect the edge state wavefunctions as much as they do for bulk states. Edge states therefore show reduced variation. Using numerical modeling including our measured values for disorder (see Supplementary), as shown in Fig.~\ref{fig:2}(c,d), we can identify these regions as the long edge and the short edge.

We next analyze the delay distribution to distinguish edge state transport from bulk states. This approach provides an unequivocal signature of localization \cite{Texier:1999,Chabanov:2001}. Fig.~\ref{fig:2}(e,f) show the measured and simulated delay distributions for the edge and bulk states in 8$\times$8 lattice sized devices and highlight the remarkable difference between edge and bulk states. For edge states, the delay distribution normalized to its average (in actuality, we used root mean square to allow for negative delay values, see Supplementary) is essentially gaussian with a gaussian width independent of system size. This behavior is characteristic of diffusive transport as seen previously in one-dimensional systems \cite{Genack:1999}. The bulk state distribution is however, asymmetric with the most probable value being less than the average. This feature is reminiscent of transport governed by localization which also has been observed earlier in the microwave regime for one-dimensional systems \cite{Chabanov:2001}. For localized transport, the delay spectrum exhibits spikes (see \ref{fig:2}(b)) which manifest in the asymmetric delay statistics. These spikes appear due to resonant tunneling through (delocalized) necklace states which are common to finite-size open systems \cite{Pendry:1987}. Therefore, even in the presence of loss, delay distribution can clearly differentiate two different regimes of transport in the same photonic system. Our measured results show a good match with numerical modeling. We observe similar behavior for other lattice sizes as well (see Supplementary).

\begin{figure}
 \centering
 \includegraphics[width=0.7\textwidth]{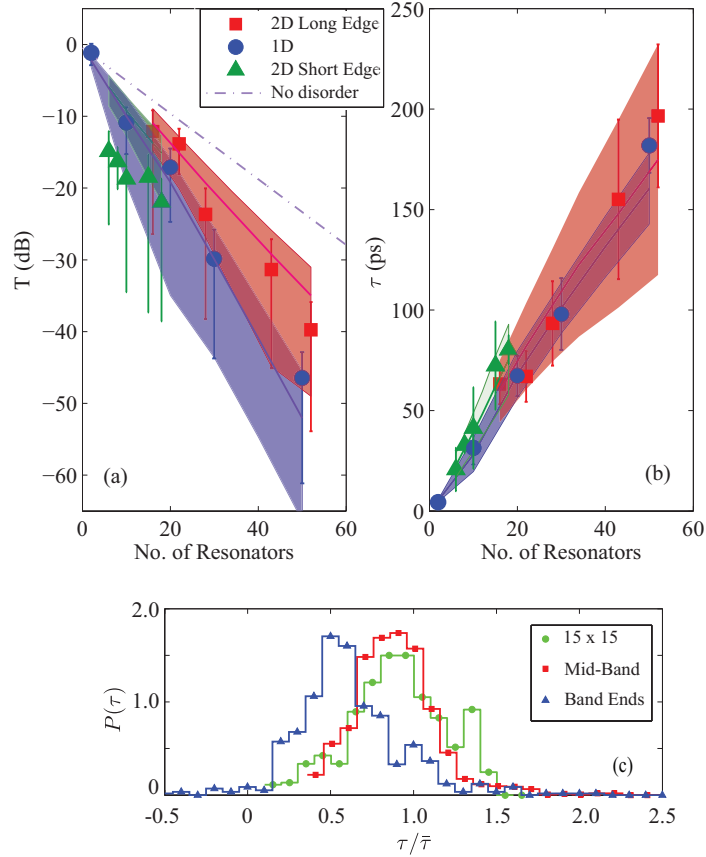}
 \caption{(a) Transmission and (b) delay-time scaling for 2D and 1D devices. Solid markers with error bars are the measured average and standard deviation (65\% confidence band) values. Solid lines with shaded areas are the simulated average and standard deviation. Also shown is the transmission when there is no disorder in the system. For 2D data, we measured (7,8,9,8,8) number of chips for (6$\times$6, 8$\times$8, 10$\times$10, 15$\times$15, 18$\times$18 ring) sized devices respectively. For the 1D data, we measured (11,15,11,12,6) number of chips for (2,10,20,30,50) ring devices. (c) Delay statistics for long edge of 15$\times$15 lattice sized devices,  mid-band and band-ends of 30-ring 1D devices. The long edge and mid-band of 1D devices show diffusive transport. The band-ends in 1D devices however show localization. }
 \label{fig:3}
\end{figure}

A test to further establish the topologically-protected nature of edge states would be a comparison of the transmission scaling with system size for an edge state against that of a topologically trivial 1D system \cite{Cooper:2010,Notomi:2008}, both with similar degrees of disorder. Fabrication-induced disorder in a 1D ring resonator array leads to a spread in the resonance wavelengths of the resonators. This impedes the forward propagation of light, increases back-reflection, i.e., less light is coupled into the array, and hence the transmission at output is reduced \cite{Ferrari:2009}. Ultimately, as the array length increases, Anderson localization halts transmission of the light \cite{Schwartz:2007,Shayan:2008,Topolancik:2007, Atlasov:2010}. Edge states, on the other hand, are unidirectional and immune to reflection caused by disorder. Therefore, transmission through edge states is expected to be less affected.

Fig. 3a shows the measured average transmission and its standard deviation across a number of chips (95 in total) for the long edge state band in 2D lattice and the mid-band of the 1D array as a function of system size, i.e. the number of resonators travelled from input to output (excluding the link resonators). Transmission in both the long-edge state as well as the 1D system decays exponentially with system size. A linear fit to measured transmission (in dB) in the long edge band gives the decay slope as -0.75(20) dB per ring but for 1D transmission the slope is -0.93(16) dB per ring, where uncertainties represent one sigma standard deviation. Transmission along long edge state can be seen to decay at a slower rate when compared to 1D transport. Simulation results using the experimentally estimated parameters are also presented in the figure. The simulated transmission decay slopes are -0.66(2) and -1.06(5) dB per ring respectively for long edge state and 1D. The experimental and simulation results are seen to agree, given that the number of devices measured for each lattice size is only $\approx$8 versus the 5000 realizations for each simulation. In order to differentiate the decay of transmission with system size resulting from resonator losses characterized by $\kappa_{in}$, from losses due to disorder - both resulting in exponential attenuation - we plot the simulated result for transmission without disorder (presented as a dashed line). In that zero disorder limit, both the 2D and 1D systems are similarly attenuated by loss with a decay rate of 0.46dB per ring. We therefore observe that disorder affects both 2D and 1D systems, but transport in edge states is less susceptible to disorder.

Fig.~\ref{fig:3}b shows the measured and simulated average delay and its standard deviation for short-edge and long-edge-state-bands. The measured delay, when plotted against the number of rings on the short and long edges of the lattice, increases linearly with a slope 3.9(9)ps and 5.4(1.0)ps per ring respectively for long and short edge states. The simulated delay slopes are 3.2(2)ps and 4.4(1)ps per ring respectively. Again, the experimental results are in agreement with the simulation. Also shown in the figure, for comparison, is the measured delay in 1D devices. That delay follows the same scaling as the edge states. However, it can be seen that the standard deviation in delay for 1D devices is less than that for edge states. This is contrary to the case of a uniform magnetic field where the standard deviation of delay in edge states remains smaller than in a 1D system \cite{Hafezi:2011}. Using simulations we have verified that this is due to the fabrication disorder of $\dPhii$ in the 2D lattice. We further compare the delay distribution for edge states and a 1D system. The normalized delay distribution for long-edge states in a 15$\times$15 lattice, and for mid-band and band-ends of a 30 ring 1D array are shown in Fig.~\ref{fig:3}c. We see that for both, edge states and the mid-band of  a 1D array, the transport is diffusive, the distribution is gaussian and the width of the distribution is independent of system size. However, the band ends of  the 1D array are localized. Using simulations, we also find that as the array length increases beyond 70 rings, even the mid-band of the 1D system shows localization.

The silicon-on-insulator technology provides a suitable platform to investigate the statistical effects of synthetic gauge fields on various transport properties and to demonstrate the localization of bulk states and robustness of edge states. Such a system could pave the way to investigate the effects of other gauge fields, including those with magnetic monopoles and floquet properties, with or without abelian features on bosonic transport. Moreover, the addition of enhanced optical nonlinearity in these ring resonator structures opens the door to intriguing questions on the nature of solutions for nonlinear transport in systems with topological order.

%%%%%%%%%%%%%%%%%%%%%%%%%%%%%%%%%%%%%%%%%%%%%%%%%%%%%%%%%%%%%%%%%%%%%%%%%%%%%%%%%%%%%%%%%%%%%%%%%%%%%%%%%%%%%%%%%%%%%%%%%%%%
% Bibliography
%%%%%%%%%%%%%%%%%%%%%%%%%%%%%%%%%%%%%%%%%%%%%%%%%%%%%%%%%%%%%%%%%%%%%%%%%%%%%%%%%%%%%%%%%%%%%%%%%%%%%%%%%%%%%%%%%%%%%%%%%%%%
\bibliographystyle{apsrev}
\bibliography{Delay_Paper_Biblio}

\textbf{Acknowledgments:} We thank A. Lobos, S. Ganeshan and J. Keutchayan for fruitful discussions. This research was supported by the U.S. Army Research Office Multidisciplinary University Research Initiative award W911NF0910406, and the NSF through the Physics Frontier Center at the Joint Quantum Institute. SF acknowledges financial support from the European Research Council (Advanced Grant SiMoSoMa).

\textbf{Disclaimer:} Certain commercial equipment, instruments or materials are identified in this paper to foster understanding. Such identification does not imply recommendation or endorsement by the National Institute of Standards and Technology, nor does it imply that the materials or equipment are necessarily the best available for the purpose.

\renewcommand{\theequation}{S\arabic{equation}}
\setcounter{equation}{0}
\renewcommand{\thefigure}{S\arabic{figure}}
\setcounter{figure}{0}

\newpage

\textbf{Supplementary Material}

\subsection{Device Fabrication and Measurements}

The devices were fabricated at the IMEC foundry using deep-UV projection photolithography \cite{Hafezi:2013}. The resonator waveguide cross-section is 510 nm in width and 220 nm in height, which allows only a single TE mode to exist in waveguides. The coupling region between all rings consists of a linear waveguide section of 7 $\mu$m with coupling gap of 180 nm, resulting in a uniform coupling rate $\J$. The system is probed using input/output waveguides coupled to the lattice, with a coupling rate $\Kex$ (Fig.~\ref{fig:1}a). Light coupled to the lattice at the input port travels through the lattice and appears at the drop port. The fraction of light which does not couple to the lattice travels to the through port. The light backscattered, due to waveguide surface roughness and reflections in the coupling region, is directed to the back-scattering port. The back-scattered light intensity is about 30dB lower than that observed on the drop port indicating negligible spin-flip disorder in the lattice. For transmission and delay-time measurements, we use an optical vector analyzer (LUNA OVA5000) based on swept wavelength interferometry \cite{Gifford:2005}. The delay-time (Wigner time) for propagation is then calculated as a derivative of phase with respect to angular frequency. Unlike transport time, Wigner time can be negative for anomalous dispersion regions around a phase jump. The negative delay values also appear in simulations where the delay is calculated similarly.

\subsection{Characterization}

To characterize the system parameters, we use one ring resonator with two coupling waveguides (add-drop filter - ADF) and one ring resonator with a single coupling waveguide (all-pass filter - APF), designed with the same parameters as those of the rings in the lattice. Transmission measurements were made on 26 single ring ADFs on different chips. Using these measurements $\Kin$, $\Kex$ and $\J$ were measured to be 2.35 GHz, 37.8 GHz and 32.0 GHz with a relative standard deviation of 20\%, 4\% and 4\% respectively. The magnetic flux $\Phii$ is designed to be  $\frac{\pi}{2}$. For 1D devices, the measured bandwidth was less than that expected from simulations with $J$=32 GHz. Therefore, for 1D simulations, a corrected value of $J$=25 GHz was used. A similar procedure was used earlier to estimate $\J$ for a 1D array of coupled resonators in \cite{Cooper:2010b}. Fabrication errors also result in a variation of the resonance frequency $\V0$ of the rings in a given lattice. The standard deviation $\dV0$ was estimated using transmission measurements on three chips with five all-pass filters each, with a physical separation of the APFs commensurate to rings in a lattice. $\dV0$ can then be used to calculate the deviation of optical path length and hence $\dPhii$ in the link rings. $\dV0$ was estimated to be 27.5 GHz and $\dPhii$ to be 0.1.

\subsection{Calibration of spectra}

Each transmission and delay spectrum shown in this work  is normalized to the corresponding measurement made away from the resonance band, at the through port. The measurements thus normalized give the actual transmission and delay incurred only through the lattice and excludes those in the coupling waveguides and connecting fibers. Because of the intrinsic spread in resonance frequencies resulting from fabrication disorders, the measured and simulated spectra have been shifted along the frequency axis to superpose them. Since the spectra are expected to be disparate in the bulk region \cite{Hafezi:2011}, we can rely only on the edge state regions to superpose them. For measured spectra, we therefore first do a manual coarse shift to align similar looking features in the expected edge state regions of the spectra. This accounts for $\dV0$ across various chips (which is much greater than $\dV0$ for a given lattice). Then we analyze the standard deviation of transmission and delay across devices as a function of frequency and find that the edge states are evident as regions with reduced noise. To verify this evidence for edge states and also to align them further, we require an algorithm based on quantitative measurements of transmission (T) and delay-time ($\tau$). Weighted delay time $W(\nu) = T(\nu) \tau(\nu)$ is one parameter that accounts for both our measurements and has been used extensively to study transport properties in random media \cite{Chabanov:2001, Tiggelen:1999}. For completely random transport, as is the case for bulk states, we expect increased variations in $T(\nu)$  and $\tau(\nu)$ and hence also in $W(\nu)$ as a function of frequency. On the contrary, transport through the edge states band follows a definite path and should therefore display regions with reduced variance in $W(\nu)$. We accordingly use the standard deviation of $W(\nu)$ to look for edge state regions and align the measured spectra. Each spectrum is shifted such that $f_{\nu}$ given by

 \be
 f_{i} = \frac{\sigma_{\nu}(T_{i}(\nu))}{\overline{T_{i}(\nu)}} + \frac{\sigma_{\nu}(\tau_{i}(\nu))}{\overline{\tau_{i}(\nu)}},
 \ee

 where $i$ refers to device index, is minimized in the designated edge state bands. The bandwidth of the long edge is found to be $\approx$10 GHz, independent of the device size while the short edge is wider (12.5 GHz -  19 GHz). For our analysis, we fix the bandwidth of the short-  and long-edge regions to be 10 GHz for all devices. For simulated spectra, we follow exactly the same protocol except for the course shift which is not required. The spectra have not been shifted along y axis and there is no re-scaling of the spectrum.

\subsection{Numerical simulations with noise}

For numerical simulations of the transmission and delay spectrum we use coupled mode analysis treating the ring resonators as lumped elements \cite{Hafezi:2011}. In the tight binding approximation, the resonators are coupled only to their nearest neighbors with a coupling rate $\J$. As shown in the supplementary section of \cite{Hafezi:2013}, $\J$ includes the response of connecting rings. The Hamiltonian of the system is characterized by the resonance frequency of each ring resonator $\W0$, its coupling rate to its nearest neighbors $\J$, the magnetic phase $\Phii$ acquired when hopping along nearest neighbors along x-axis and $\Kex$, the coupling rate to probe waveguide.

To include lattice disorder into this Hamiltonian, we impose random variations on each of the parameters with a gaussian probability distribution around the mean. For each numerical realization of the lattice, each resonator ring has noise added to it resonance frequency, coupling rate to neighboring resonators and also to the magnetic phase acquired in hopping along x-axis. To get mean transmission and delay, we then average the results over 5000 realizations for each device type. Required simulation parameters and their deviations, characterizing the system and its disorder, have been measured using multiple add-drop filters as described above.

The fact that the measured transmission scales exponentially and the delay scales linearly with the number of rings on the edge of the lattice, and both match well with the simulated results, reinforce our claim that the low noise areas are in fact the short and long edge regions. From simulation, we found that the main disorder terms affecting transmission in a 2D lattice are $\dW0$ and variations in the otherwise uniform magnetic field i.e. $\dPhii$, whereas for a 1D array only the first term is applicable since there is no magnetic field. In the absence of $\dPhii$, the transmission in the edge state would be even closer to the dashed line with no disorder. The short-edge transmission in our system was however consistently found to be much lower ($\approx$8 dB for 6x6 devices) than expected using simulations, but it tends to match simulation results for bigger sized devices. We expect this to be the result of some systematic problem with our fabrication process which couples less light to the short edge at the input port and hence produces a non-zero intercept on the transmission axis. Using through port data, we verified the reduction in coupling efficiency for the short edge band, for all devices. \\

\subsection{Spectrum and delay statistics for 15$\times$15-lattice-sized devices}

\begin{figure}
\centering
\includegraphics[width=0.8\textwidth]{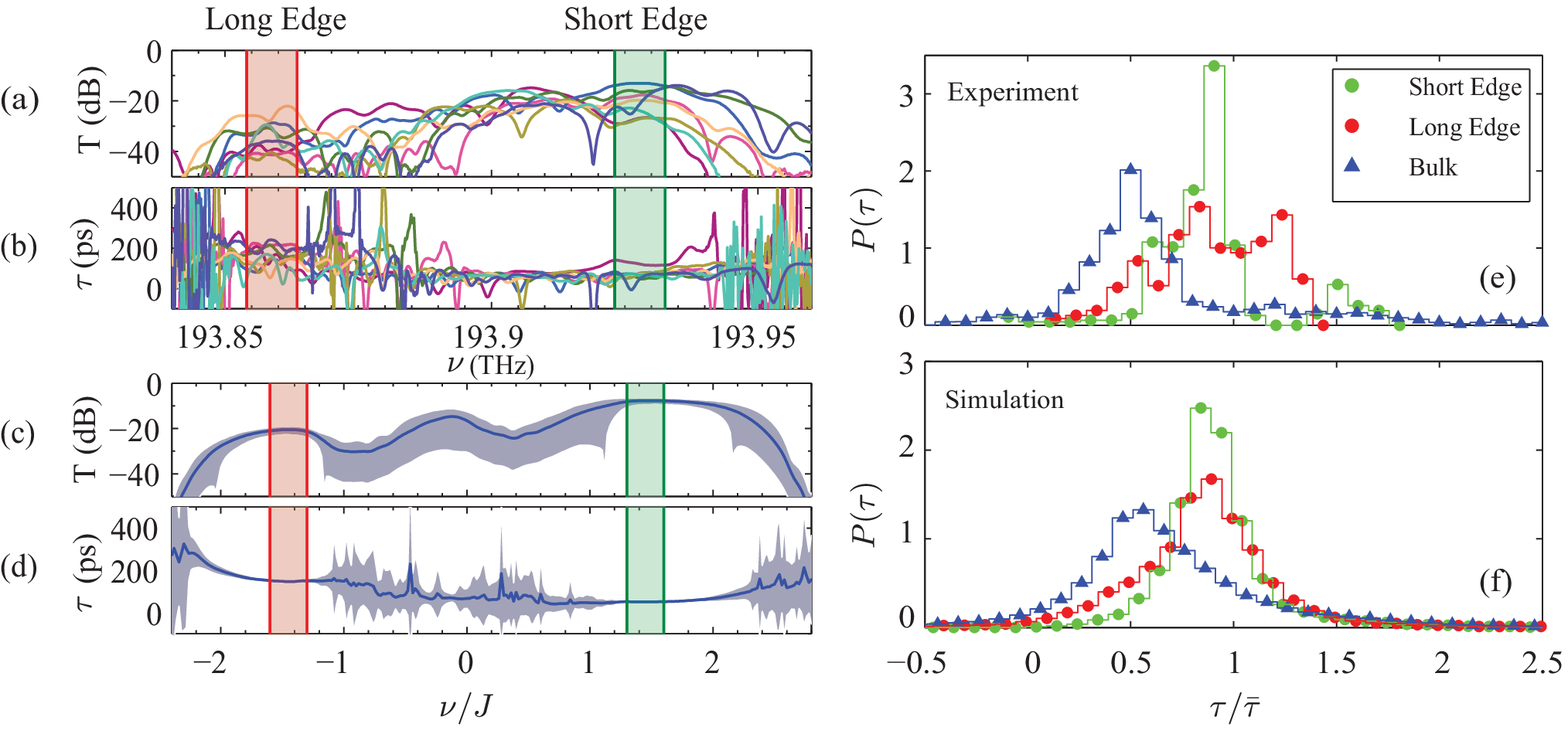}
\caption{(a,b) Measured and (c,d) simulated transmission and delay spectra for 15$\times$15-lattice-sized devices. (e,f) Measured and simulated delay distribution. The data is taken across 8 devices.}
\label{fig:4}
\end{figure}

Fig.~\ref{fig:4} shows the measured and simulated transmission and delay statistics for 15$\times$15-lattice-sized devices. As was seen in Fig.~2, the transmission and delay spectra show two regions with reduced noise. The delay statistics is also similar to what is observed for 8$\times$8-lattice-sized devices. Edge state transport is diffusive whereas the bulk state is localized.

\subsection{Gaussian fits to delay distribution}

\begin{figure}[h]
\centering
\includegraphics[width=0.8\textwidth]{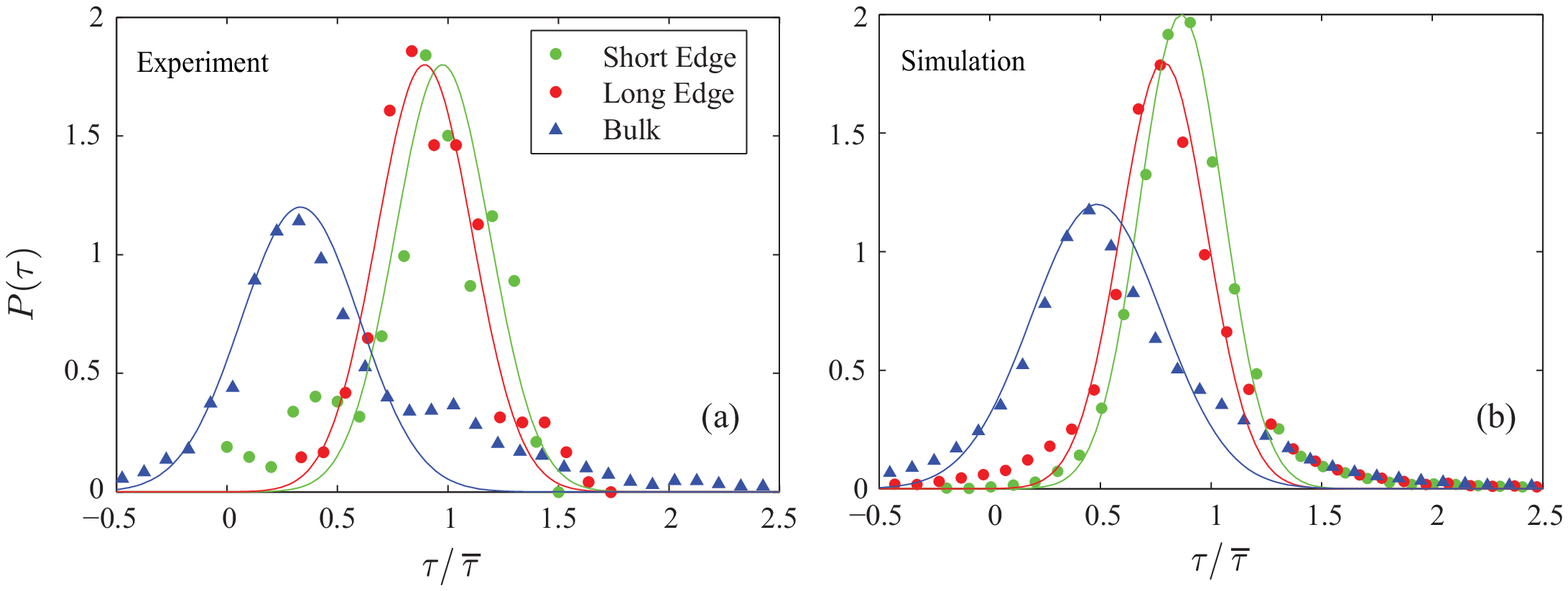}
\caption{(a) Gaussian fits to measured delay distributions for edge and bulk states in 8$\times$8-lattice-sized devices. The bulk state distribution shows a systematic deviation from the Gaussian fit for longer delay values. (b) Gaussian fits to simulated delay distribution, showing similar behavior for bulk states.}
\label{fig:5}
\end{figure}

Fig.~\ref{fig:5} plots Gaussian fits to measured and simulated delay distributions for 8$\times$8-lattice-sized devices. The bulk states deviate systematically from the Gaussian fit, towards longer delay times.

\subsection{Localization in 2D Lattice}

\begin{figure}[h]
\centering
\includegraphics[width=0.45\textwidth]{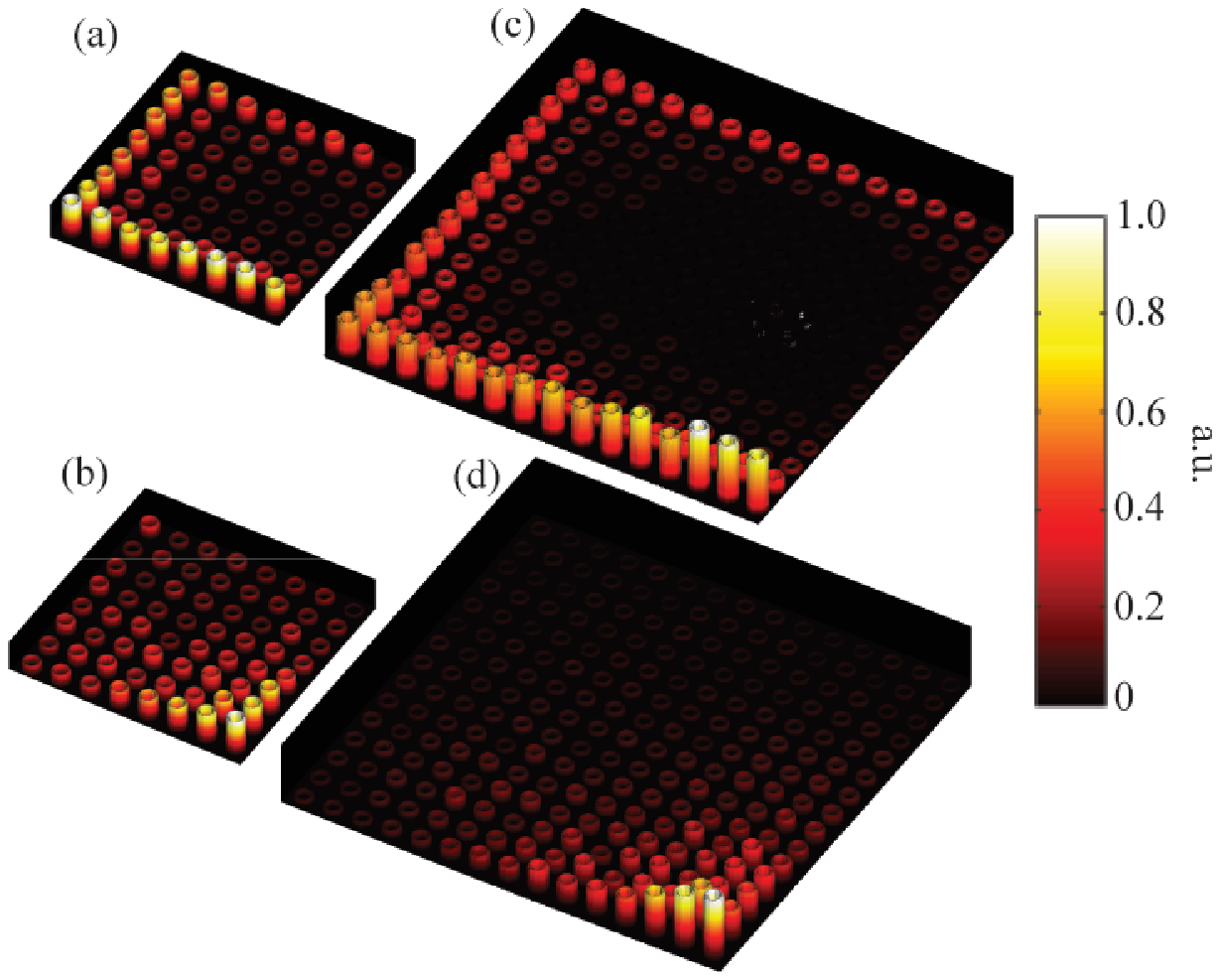}
\caption{(a,b) Simulated intensity plots for long edge and bulk states respectively for 8$\times$8 lattice. The bulk states are localized near the input whereas the edge states are extended. (c,d) Simulated intensity plot for 15$\times$15 lattice. The edge states are still extended and the bulk localized.}
\label{fig:intensity}
\end{figure}

Fig.~\ref{fig:intensity} (a,b) shows the simulated long edge and bulk state intensity for an 8$\times$8 lattice, with disorder, averaged over 50 realizations. In the presence of disorder, the long edge state spans the complete edge of the lattice, whereas the bulk states are localized near the input port. As the system size increases, as shown in Fig.~\ref{fig:intensity}(c,d), the edge state wavefunction still extends across the lattice edge, but the bulk state is again localized with localization extent independent of lattice size.

\subsection{Localization in a 1D array}

\begin{figure}[h]
\centering
\includegraphics[width=0.45\textwidth]{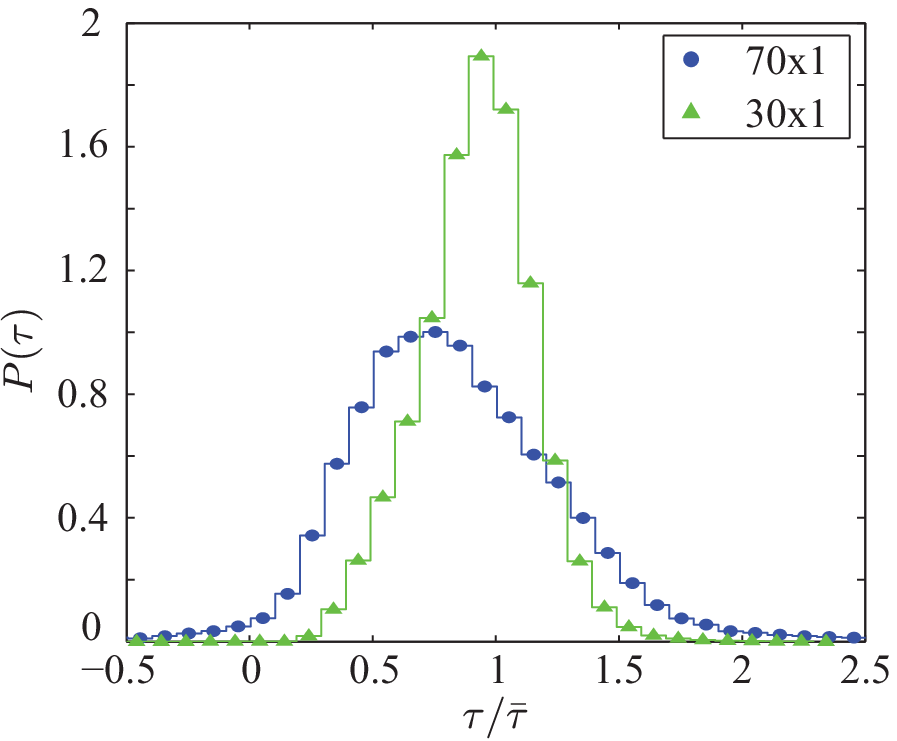}
\caption{Simulated delay statistics for a 1D array with 70 rings shows localization.}
\label{fig:7}
\end{figure}

Fig.~\ref{fig:7} shows the simulated delay statistics for a 1D array with 70 rings. The probability distribution is asymmetric indicating the onset of localization.

\subsection{Measured spectrum for 1D devices with 10 resonator rings}

Fig.~S5 shows the measured and simulated transmission and delay spectrum for eleven, 1D devices with 10 resonator rings. The device to device variations in transmission and delay are more or less independent of frequency.

\begin{figure}[h]
\centering
\includegraphics[width=0.8\textwidth]{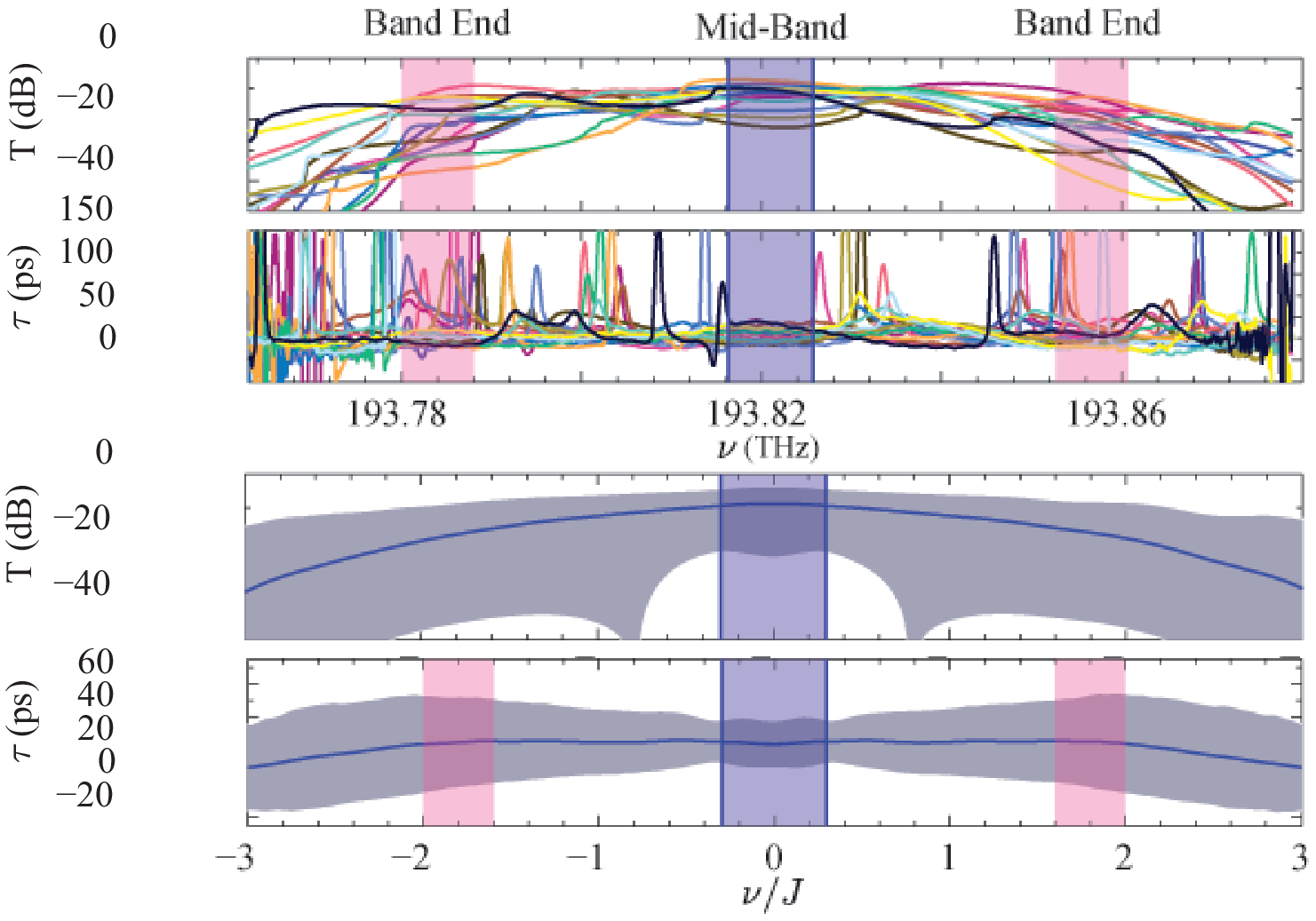}
\caption{Measured and simulated transmission and delay spectrum for a 1D system with 10 rings}
\label{fig:8}
\end{figure}

\subsection{Device yield}

Table I shows the detailed device yield. For 2D devices, the device yield was found to be $\approx$23\%. For 1D devices (other than 50 rings), the device yield was 100\%. For 50 ring devices, the yield was $\approx$50\%. The devices with a very noisy or attenuated spectrum were considered as bad.

\newcolumntype{C}[1]{>{\centering}p{#1}}
\begin{table}[h]
\label{fig:T1}
\centering
%\begin{tabular}{| c | c | c | c | c |}
\begin{tabular}{| C{2.5cm} | C{2.5cm} | C{2.5cm} | C{2.5cm} | c |}
  \hline
 \textbf{\textcolor{blue}{Device size}}  & \textbf{\textcolor{blue}{Worked}} & \textbf{\textcolor{blue}{Bad}} & \textbf{\textcolor{blue}{Did not scan}} & \textbf{\textcolor{blue}{~~~~~~Total~~~~~~}}\\
  \hline\hline
  \textbf{\textcolor{blue}{6$\times$6}}   & 7  & 28 & 0  & 35  \\
  \textbf{\textcolor{blue}{8$\times$8}}   & 8  & 27 & 0  & 35  \\
  \textbf{\textcolor{blue}{10$\times$10}} & 9  & 26 & 0  & 35  \\
  \textbf{\textcolor{blue}{15$\times$15}} & 8  & 27 & 0  & 35  \\
  \textbf{\textcolor{blue}{18$\times$18}} & 8  & 27 & 0  & 35  \\
  \textbf{\textcolor{blue}{2$\times$1}}   & 11 & 0  & 21 & 35  \\
  \textbf{\textcolor{blue}{10$\times$1}}  & 15 & 0  & 20 & 35  \\
  \textbf{\textcolor{blue}{20$\times$1}}  & 11 & 0  & 24 & 35  \\
  \textbf{\textcolor{blue}{30$\times$1}}  & 12 & 0  & 23 & 35  \\
  \textbf{\textcolor{blue}{50$\times$1}}  & 6  & 7  & 22 & 35  \\
\hline
\end{tabular}

\caption{Number of devices measured, good and bad, for each device type}
\end{table}

\end{document}